# A GPU Tool for Efficient, Accurate, and Realistic Simulation of Cone Beam CT Projections

Xun Jia[1], Hao Yan[1], Laura Cerviño[1], Michael Folkerts[1,2], and Steve B. Jiang[1]

[1]Center for Advanced Radiotherapy Technologies and Department of Radiation Medicine and Applied Sciences, University of California San Diego, La Jolla, CA 92037, USA

[2]Department of Physics, University of California San Diego, La Jolla, CA 92093, USA

E-mail: sbjiang@ucsd.edu, xunjia@ucsd.edu

**Purpose:** Simulation of x-ray projection images plays an important role in cone beam CT (CBCT) related research projects, such as the design of reconstruction algorithms or scanners. A projection image contains primary signal, scatter signal, and noise. It is computationally demanding to perform accurate and realistic computations for all of these components. In this work, we develop a package on GPU, called gDRR, for the accurate and efficient computations of x-ray projection images in CBCT under clinically realistic conditions.

**Methods:** The primary signal is computed by a tri-linear ray-tracing algorithm. A Monte Carlo (MC) simulation is then performed, yielding the primary signal and the scatter signal, both with noise. A denoising process specifically designed for Poisson noise removal is applied to obtain a smooth scatter signal. The noise component is then obtained by combining the difference between the MC primary and the ray-tracing primary signals, and the difference between the MC simulated scatter and the denoised scatter signals. Finally, a calibration step converts the calculated noise signal into a realistic one by scaling its amplitude according to a specified mAs level. The computations of gDRR include a number of realistic features, e.g. a bowtie filter, a poly-energetic spectrum, and detector response. The implementation is fine tuned for a GPU platform to yield high computational efficiency.

**Results:** For a typical CBCT projection with a poly-energetic spectrum, the calculation time for the primary signal using the ray-tracing algorithms is 1.2~2.3 sec, while the MC simulations take 28.1~95.3 sec, depending on the voxel size. Computation time for all other steps is negligible. The ray-tracing primary signal matches well with the primary part of the MC simulation result. The MC simulated scatter signal using gDRR is in agreement with EGSnrc results with a relative difference of 3.8%. A noise calibration process is conducted to calibrate gDRR against a real CBCT scanner. The calculated projections are accurate and realistic, such that beam-hardening artifacts and scatter artifacts can be reproduced using the simulated projections. The noise amplitudes in the CBCT images reconstructed from the simulated projections also agree with those in the measured images at corresponding mAs levels.

**Conclusions:** A GPU computational tool, gDRR, has been developed for the accurate and efficient simulations of x-ray projections of CBCT with realistic configurations.

# 1   Introduction

Cone beam computed tomography (CBCT)[1,2] has become an important tool in medical imaging for direct visualization of patient anatomy. In many CBCT-related research topics, for instance the design of CBCT scanners and the development of reconstruction algorithms, it is highly desirable to perform accurate and realistic simulations to obtain x-ray projection images. Not only is this a cost-effective way of acquiring data without performing real experiments, it also offers the opportunities and freedoms to disentangle all the physical effects in CBCT, such as various types of scatter signals, so that researchers can specifically focus their studies.

Generally speaking, there are three components that one needs to consider in a projection image, namely, primary signal, scatter signal, and noise signal, all of which are of interest to certain research projects and applications. The computations of these components are very demanding, especially if one would like to achieve a high level of accuracy and realism. Over the years, there have been a number of research efforts dedicated to the computations of these components.

The primary signal characterizes x-ray attenuation while traveling from an x-ray source to a detector pixel. This signal forms the fundamentals for the CT technology. Therefore, computation of the primary signal is widely employed in studies regarding the design and validation of reconstruction algorithms. Although it is conceptually straightforward to compute this signal by ray-tracing methods[3-5], it becomes very computationally intensive to obtain accurate results in a realistic context. For instance, repeated ray-tracing calculations are needed in those cases with a poly-energetic x-ray spectrum, each corresponding to an energy channel.

Scatter signal is the second component of interest. As it is the primary contamination in CBCT imaging, calculating this signal serves as the basis for understanding, modeling, and removing scatter[6-8]. Monte Carlo (MC) methods have been widely used for scatter calculations[9-12] due to its faithful descriptions of the underlying physical process and the accurate considerations of the problem geometry. Nonetheless, the extremely prolonged computation time required to achieve an acceptable precision level has seriously impeded its applications. To speed up the computations, variance reduction techniques have been utilized[10,11]. Moreover, because of the smoothness of a scatter signal, it has also been proposed to compute it on a detector grid with a low resolution and a large pixel size to improve signal-to-noise ratio, and hence effectively reduce computational time[12,13]. Yet, the computation time, especially when computing a large number of projections, is still not satisfactory.

The third component in a projection image is noise. Since it is usually desirable to acquire CBCT projections at low mAs levels for the consideration of imaging dose reduction, studying properties of the amplified noise is necessary to facilitate the development of noise removal techniques. In the past, a number of noise models have been proposed[14-16], where variance of the noise at each pixel is usually assumed to be a function of the primary signal and the parameters in these models are obtained by fitting against measurement data. However, these models generate noise signals only in a phenomenological manner, and the physical process of noise formation, namely the quantum fluctuation of photons arrival at a detector pixel, is neglected.

    To our knowledge, there is no single package that computes all of these components together to an adequate degree in terms of combined accuracy, realism, and efficiency. This fact motivates us to develop a new package, gDRR, aiming at meeting all of these requirements. Generally, satisfactory accuracy and realism usually lead to compromised efficiency. gDRR overcomes this problem by employing the high-performance platform of graphics processing unit (GPU), as well as simulation algorithms and schemes suitable for GPU. Recently, GPUs have been increasingly utilized in medical physics to speed up computationally intensive tasks[17-26]. In particular, it has been demonstrated that GPUs can greatly enhance the MC simulation efficiency of particle transport[27-32], the most computationally demanding task in x-ray projection simulations. Among them, MC-GPU[30, 31] has been developed for x-ray radiograph simulations, and up to ~30 times speed-up has been observed compared to CPU simulations. Yet, the photon transport functions in MC-GPU are essentially a straightforward translation of the PENELOPE[33] subroutines and the well tuned implementations in PENELOPE for CPU may not be optimal on the GPU platform. Moreover, a number of realistic features required in simulations are missing in MC-GPU, such as detector response.

    In this work, we will present our recent progress towards a high performance x-ray imaging simulation package, gDRR. This package computes primary signals using ray-tracing algorithms, while an MC simulation optimized for the GPU platform is employed to obtain the primary and the scatter signals with noise. A denoising procedure designed for Poisson noise removal is utilized to yield the scatter signal that is smooth across the detector. Finally, gDRR computes the difference between the simulation results with and without noise, yielding the computed noise components, whose amplitude is then properly scaled according to a specified mAs level. Realistic CBCT geometry and detailed physical aspects are considered in gDRR. The entire computation is performed on GPU, which leads to a high efficiency.

## 2  Methods and Materials

### 2.1  *System setup*

Let us consider the geometry for a CBCT system as illustrated in Figure 1. An x-ray source is at one side of a patient, which is able to rotate inside the *xOy* plane about the *z* axis. The location of the x-ray source *S* is parameterized by the source-to-axis distance (SAD) and the rotation angle $\varphi$ between $\overrightarrow{SO}$ and the positive *x* direction. An x-ray image detector is perpendicular to the source rotational plane *xOy* as well as the direction $\overrightarrow{OP}$. The imager location is defined by the axis-to-imager distance (AID) and another angle $\theta$ between $\overrightarrow{SO}$ and $\overrightarrow{OP}$. This configuration allows for an easy placement of the imager not necessarily in the opposite side of the source required in studies such as Compton scatter tomography[34]. A coordinate system $(u, v)$ is defined on the detector plane with its origin at the point $P$ and the $v$ axis is parallel to the $z$ axis.

gDRR computes the projections of a voxelized phantom represented by a 3-d array indicated by the cube shown in Figure 1. At each voxel, a material type index $i(\boldsymbol{x})$ and a density value $\rho(\boldsymbol{x})$ are specified. X-ray mass attenuation coefficient $\hat{\mu}_k(i, E)$ is also available corresponding to each type of material $i$, where the subscript $k = 1,2,3$ labels the three interaction types relevant in the kilo-voltage energy regime, namely Rayleigh scattering, Compton scattering, and photoelectric effect.

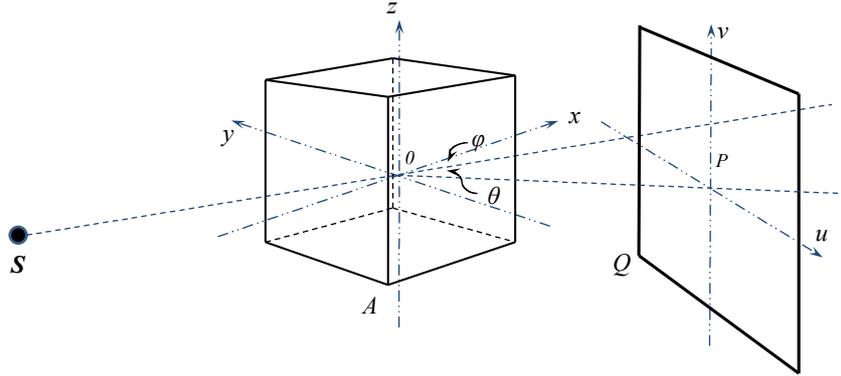

Figure 1. An illustration of simulation geometry in gDRR.

As for the image detector, it is modeled to be a 2-d pixel array. Each detector pixel acquires photon signals in energy integration mode, where the total photon energy deposited to the pixel is recorded. Detector response is considered in gDRR through a user supplied response curve $r(E)$, which specifies the amount of energy deposited by an incoming photon of energy $E$. An example of the detector response curve is shown in Figure 2(a). In principle, the detector response could also be pixel dependent. In our calculation, we ignore this pixel dependence for simplicity.

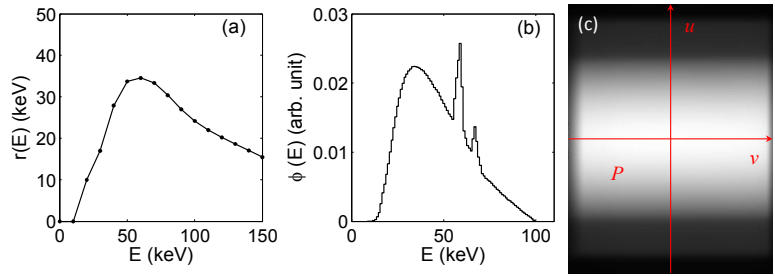

Figure 2. From (a) to (c): a typical detector response curve, a typical 100kVp source energy spectrum, and a photon fluence map after a full-fan bowtie filter.

gDRR does not transport photons inside an x-ray source, *e.g.* the x-ray target and the bowtie filter. Therefore, all quantities used to characterize the source properties are defined after the bowtie filter. Specifically, the x-ray source is defined by its energy spectrum and its fluence map. The energy spectrum $\phi(E)$ describes the probability density of a source photon as a function of its energy $E$. A typical 100 kVp energy spectrum is depicted in Figure 2(b). User can specify such an energy spectrum by using the method developed by Boone *et al.*[35]. As for the photon fluence map, $w(\boldsymbol{u})$, it is used to specify the probability density of a photon traveling towards the detector coordinate $\boldsymbol{u} = (u, v)$ after coming from the source and can be obtained by acquiring a CBCT air scan image. A typical example of the fluence map after a full-fan bowtie filter is illustrated in Figure 2(c). In gDRR, both the spectrum and the fluence map are normalized, such that $\int dE\, \phi(E) = 1$ and $\int d\boldsymbol{u}\, w(\boldsymbol{u}) = 1$.

## 2.2 Overall computational structure

Figure 3 illustrates the overall workflow and data flow in gDRR. Those key steps are labeled with numbers, while the shaded boxes are the data sets generated during calculations. Let us denote the primary, scatter, and noise signals by $P(\boldsymbol{u})$, $S(\boldsymbol{u})$, and $N(\boldsymbol{u})$. In gDRR, Step 1 utilizes a ray-tracing algorithm to compute the primary x-ray attenuation signal at the detector $P(\boldsymbol{u})$. Step 2 calculates the primary signal $\hat{P}(\boldsymbol{u})$ and the scatter signal $\hat{S}(\boldsymbol{u})$ via MC simulations and the results contains noise due to the stochastic nature of the MC method. The difference between the noisy primary signal from the MC simulation and the noise-free one from the ray-tracing method yields the noise signal in the primary signal calculation $\widehat{N}_P(\boldsymbol{u}) = \hat{P}(\boldsymbol{u}) - P(\boldsymbol{u})$. A denoising technique (Step 3) is then applied to the noisy scatter signals $\hat{S}(\boldsymbol{u})$, leading to the smoothed scatter signal $S(\boldsymbol{u})$, as well as the noise on the scatter part $\widehat{N}_S(\boldsymbol{u}) = \hat{S}(\boldsymbol{u}) - S(\boldsymbol{u})$. The noise from the primary and that from the scatter add up to the total simulated noise in the projection $\widehat{N}(\boldsymbol{u}) = \widehat{N}_P(\boldsymbol{u}) + \widehat{N}_S(\boldsymbol{u})$. Finally, a noise scaling step in Step 4 is invoked to scale the noise amplitude according to an mAs level specified in the simulation, resulting in the final noise signal $N(\boldsymbol{u}) = \alpha \widehat{N}(\boldsymbol{u})$. After launching gDRR, three components in a CBCT projection are computed, namely the primary signal $P(\boldsymbol{u})$, the scatter signal $S(\boldsymbol{u})$, and the noise signal $N(\boldsymbol{u})$. The following subsections will be devoted to the description of the detailed computational strategies in those key steps.

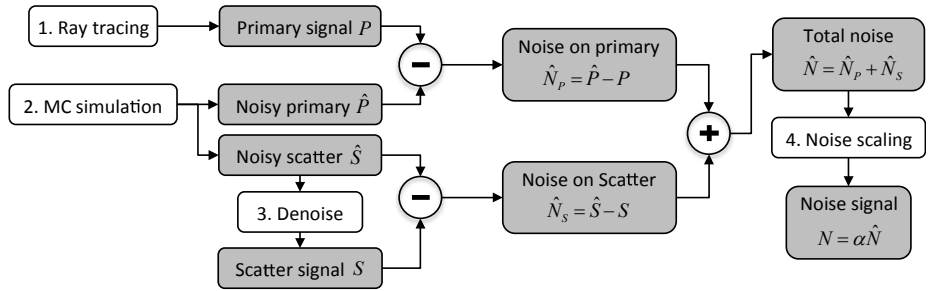

Figure 3. Task and data flow of gDRR. Boxes with numbers are those key steps in gDRR, while shaded boxes indicate key data sets generated during calculations.

## 2.3 Ray-tracing for primary signal calculation

The primary signal at a detector pixel corresponds to the x-ray attenuation process while photons travel from the source to the detector pixel. This process can be accurately modeled by the Beer-Lambert law. In the context with a poly-energetic source spectrum, fluence map, and detector response, the primary x-ray signal at a pixel $\boldsymbol{u}$ is expressed as

$$P(\boldsymbol{u}) = \int dE\, \phi(E) w(\boldsymbol{u}) r(E) \exp\left[-\int_L dl\, \mu(\boldsymbol{x}, E)\right]. \tag{1}$$

Inside the exponential term, a line integral represents the radiological length for an energy $E$, where the path $L$ is a straight line connecting the source to the detector pixel at $\boldsymbol{u}$ and $\mu(\boldsymbol{x}, E) = \rho(\boldsymbol{x}) \sum_k \hat{\mu}_k(i(\boldsymbol{x}), E)$ is the total x-ray linear attenuation coefficient at a spatial location $\boldsymbol{x}$ for the energy $E$.

In gDRR, the first integration over energy is approximated by a discrete summation over all energy channels considered. Within each energy channel, the evaluation of the line integral for the corresponding radiological length is needed. This line integral is usually evaluated using Siddon's ray-tracing algorithm[3]. However, it is known that the Siddon's algorithm leads to square-block like artifacts in the projection image, especially when the voxel size is large. In gDRR, we utilize a tri-linear interpolation algorithm to generate more realistic projections. Specifically, we divide the x-ray path into a set of intervals of equal length $\Delta l$ labeled by $j$ and compute the linear attenuation $\mu(\boldsymbol{x}_j, E)$ at the midpoint of each interval using a tri-linear interpolation scheme. The sum over all intervals $\sum_j \Delta l\, \mu(\boldsymbol{x}_j, E)$ is considered as the radiological length. Mathematically, it can be proven that the numerical result converges to the line integral $\int_L dl\, \mu(\boldsymbol{x}, E)$ in the limit of zero voxel size and $\Delta l \to 0$. To avoid over smoothing caused by a large step size $\Delta l$, gDRR sets $\Delta l$ to be half of the voxel size, which have been shown to be sufficiently small in our calculations.

In terms of computation, it is straightforward to implement the algorithm on a GPU platform. By simply having each GPU thread compute the projection value at one pixel location $\boldsymbol{u}$, considerable speed-up factors can be obtained due to the vastly available GPU threads. In our implementation, GPU texture memory is used to store the voxel linear attenuation coefficients to enable fast memory access. Hardware-supported linear interpolation is also used in the tri-linear interpolation algorithm.

*2.4    Monte Carlo simulation*

A GPU-based MC simulation is also developed for photon transport in the energy range from 1 keV to 150 keV. Specifically, multiple GPU threads are launched to transport a group of photons simultaneously, with one thread tracking a photon history. Within each thread, a source photon is first generated at the x-ray source according to the user specified energy spectrum and fluence map. The photon transport process is then handled by using the Woodcock tracking method[36], which significantly increases the simulation efficiency of voxel boundary crossing process. Three possible physical interactions are considered for photons in this energy range, namely Compton, Rayleigh, and photoelectric absorption. In an event of photoelectric effect, the photon transport process is terminated. After Compton or Rayleigh scattering events, the scattering angles are sampled according to corresponding differential cross section formula using the techniques developed in gCTD[37], a package for fast patient-specific CT/CBCT dose calculations using MC method. The photon is tracked until it is absorbed or it escapes from the phantom. This process is repeatedly performed till a preset number of photon histories are simulated.

Two counters are designed to store the primary signal $\hat{P}(\boldsymbol{u})$ and the scatter signal $\hat{S}(\boldsymbol{u})$ at the image detector, respectively. Meanwhile, an indicator is carried by each photon that records if any scattering events have taken place during the transport. In the case when a photon exits from the phantom, a simple geometrical calculation determines if it hits the detector. If so, an amount of energy $r(E)$ is recorded at a corresponding detector pixel in either the primary counter or the scatter counter depending on whether some scatter events have occurred, where $E$ is the photon energy before hitting the detector and $r(E)$ is the response curve. The user also has the option to tally the signal of a specific type, such as the first order Compton.

One issue in the MC simulation is that the way of density interpolation may impact on the projection image quality. In an MC simulation, a voxelized phantom image is defined by the user. Conventionally, it is assumed that each voxel is homogeneous, as no further information is given regarding the variations of material properties at a sub-voxel length scale. Yet, akin to the aforementioned Siddon's ray-tracing algorithm for primary signal calculation, such a configuration leads to an apparent artifact in the simulated primary image due to the finite voxel size. This artifact makes the primary signal obtained from the MC simulation not compatible with the one from the tri-linear interpolation algorithm, when it comes to the noise calibration step to be discussed later. To overcome this problem, we employed a tri-linear interpolation strategy on the density grid used in the MC simulation. As such, whenever a density value is requested by a photon, it calculates the value at the photon's instant location using the tri-linear interpolation scheme. The underlying assumption is that the density of a phantom smoothly varies in the 3-dimensional space. Such a strategy effectively eliminates the finite voxel size artifact in the MC-simulated primary projection images. As for the scatter signals, the impacts of this density interpolation scheme seems to be minimal, as the scattered photons towards various directions smear out this effect.

## 2.5   *Noise removal in scatter signals*

Due to the randomness inherent to the MC method, noise exists in both the scatter and the primary signals. For the scatter component, it is expected that its ground truth signal varies smoothly along the spatial dimension. This assumption allows us to perform some noise removal techniques to retrieve the true scatter signal from the noise contaminated one obtained from the MC simulations. A powerful noise removal algorithm also enables us to estimate the scatter signals based on MC simulations with much reduced number of photons, improving computational efficiency. In gDRR, we develop an effective method for this purpose by solving an optimization problem. We assume the noise signal $\hat{S}(\boldsymbol{u})$ at a pixel in MC simulations follows a Poisson distribution with an underlying true signal $S(\boldsymbol{u})$, which is determined by solving such an optimization problem

$$S(\boldsymbol{u}) = \text{argmin}_S E[S] = \text{argmin}_S \int d\boldsymbol{u} \left( S - \hat{S} \log S \right) + \frac{\beta}{2} \int d\boldsymbol{u} |\nabla S|^2. \tag{2}$$

There are two terms in the energy function $E[S]$. The first one is a data-fidelity term that is customized for Poisson noise[38], while the second one is a penalty term that ensures the smoothness of the recovered solution $S(\boldsymbol{u})$. $\beta$ is a constant to adjust the relative weights between the two terms. Such an energy function is convex, and hence it is sufficient to consider the optimality condition to be satisfied by the solution, namely

$$0 = \frac{\delta E}{\delta S} = \left(1 - \frac{\hat{S}}{S}\right) - \beta \nabla^2 S. \tag{3}$$

After discretizing the Laplacian operator $\nabla^2$ using a standard numerical scheme, we arrive at

$$0 = \left[1 - \frac{\hat{S}(i,j)}{S(i,j)}\right] - \beta[\Sigma S(i,j) - 4S(i,j)], \tag{4}$$

where $i$ and $j$ are pixel location indices on the detector array and $\Sigma S(i,j)$ is a short notation for $S(i+1,j) + S(i,j+1) + S(i-1,j) + S(i,j-1)$. With this numerical scheme, we can rearrange this equation and design a successive over-relaxation algorithm[39] that solves the optimization problem by iterating

$$S^{(k+1)}(i,j) = (1-\omega)S^{(k)}(i,j) + \frac{\omega}{4}[\Sigma S^{(k)}(i,j) - \frac{1}{\beta}(1 - \frac{\hat{S}(i,j)}{S^{(k)}(i,j)})], \tag{5}$$

where the superscript $k$ is an index for the iteration step. The parameter $\omega$ and the first term in this equation are introduced to speed up convergence. In practice, an empirical value of $\omega = 0.8$ is used in our implementation. Although the solution is expected to be independent of the initial guess $S^{(0)}$ due to the convex nature of this problem, it is found that the choice of $S^{(0)} = \hat{S}$ leads to faster convergence than other initialization values we considered. Such a denoising technique is particularly suitable for GPU-based parallel computation, as the component-wise multiplication and division in Eq. (5) can be easily carried out by GPU.

*2.6   Noise calibration*

After the above calculations, the noise component on a projection image can be estimated by combining that from the primary signal and that from the scatter signal, namely

$$\widehat{N} = \widehat{N}_P + \widehat{N}_S = (\hat{P} - P) + (\hat{S} - S). \tag{4}$$

Because the formation of this noise component mimics the realistic noise generation in a CBCT scan, *i.e.*, the quantum fluctuation of photons hitting a detector, the physical properties of the noise signal obtained as such are expected to be accurate given that a large enough number of photons are simulated in the MC calculations. Nonetheless, the noise amplitude in a real scan depends on the actual number of photons emitted by the x-ray source $n_{\text{act}}$, while the noise amplitude in $\widehat{N}$ is governed by the number of source photons in the simulation $n_{\text{sim}}$, which is typically much less than $n_{\text{act}}$. Therefore, it is necessary to scale the noise amplitude to yield a correct level of noise.

In our simulation, the primary signal given in Eq. (1) is expressed in terms of per particle and all the MC simulation results are normalized by the number of source photons. For the noise signal obtained as such, it can be expected that the noise amplitude at a detector pixel is approximately proportional to $1/\sqrt{n}$, where $n$ is the number of photons hitting the detector pixel. While the exact value of $n$ is unknown for both the simulation study and real experiments, it is reasonable to expect that it is proportional to $n_{\text{act}}$ in an experiment and to $n_{\text{sim}}$ in the MC simulation. Furthermore, $n_{\text{act}}$ is linearly related to the mAs level, $I$, used in an experiment. In consideration all of these factors, we propose to scale the calculated noise $\widehat{N}$ as

$$N = \alpha \widehat{N} = \widehat{N}\sqrt{\frac{n_{\text{sim}}}{\zeta I}}, \quad (5)$$

where $I$ is the mAs level and $\zeta$ is an unknown factor that can be interpreted as the effective number of source photons at unitary mAs from an x-ray source in an experiment. The exact value of $\zeta$ apparently depends on the specific CBCT machine used and can be determined by a calibration process as described in the following.

In principle, the calibration process can be accomplished by equating the noise amplitude of the calculated projection data with that of the measurements. Let us first denote the noise amplitude at a pixel $\boldsymbol{u}$ on an x-ray projection of a calibration phantom as $\sigma_X(\boldsymbol{u}, I)$ in an experiment with an mAs level of $I$. Meanwhile, we can obtain the calculated noise amplitude in gDRR denoted by $\sigma_{\widehat{N}}(\boldsymbol{u})$. If our assumption regarding the noise model holds, it follows that

$$\sigma_X(\boldsymbol{u}, I) = \sqrt{\frac{n_{\text{sim}}}{\zeta I}} \sigma_{\widehat{N}}(\boldsymbol{u}), \quad (6)$$

*i.e.*, the function $\sigma_X(\boldsymbol{u}, I)/\sigma_{\widehat{N}}(\boldsymbol{u})$ should be a constant value that is independent of the coordinate $\boldsymbol{u}$ and only depends on the mAs level $I$ for a given number of photons in the simulation. The level of this constant linearly decreases as $1/\sqrt{I}$, as the mAs level increases, and the slope of this linear relationship indicate the level of $\zeta$.

As such, let us take the calibration of gDRR against a kV on-board imaging (OBI) system integrated in a TrueBeam medical linear accelerator (Varian Medical System, Palo Alto, CA, USA) as an example. We have acquired CBCT scans of a Catphan®600 phantom (The Phantom Laboratory, Inc., Salem, NY, USA) under various mAs levels. We specifically focus on the regions on the projection image corresponding to the homogeneous phantom layer. For a fixed coordinate $\boldsymbol{u}$ inside this region, the standard deviation $\sigma_X(\boldsymbol{u}, I)$ can be estimated by using the pixel values at this coordinate in the projections at different angles. The underlying assumption is that the projection geometry and the phantom in this layer is approximately rotationally symmetric, and pixel values at a fixed coordinate in different projections can be interpreted as results from different experimental realizations. Meanwhile, the Catphan phantom is digitized based on its CT image and the projection images are calculated using gDRR. The standard deviation $\sigma_{\widehat{N}}(\boldsymbol{u})$ can be determined in the same manner. For each mAs level of the experimentally acquired data, we compute $\sigma_X(\boldsymbol{u}, I)/\sigma_{\hat{N}}(\boldsymbol{u})$ and plotted this value as a function of $\boldsymbol{u}$. The resulting constant, independent of $\boldsymbol{u}$, serves as a test regarding the validity of the noise model in Eq. (5). Finally, we plot the spatial average value of $\sigma_X(\boldsymbol{u}, I)/\sigma_{\hat{N}}(\boldsymbol{u})$ as a function of $1/\sqrt{I}$ and the data are found to be on a straight line. A linear regression yields the slope $k$ of this line and hence the level of $\zeta$ can be derived as $\zeta = n_{\text{sim}}/k^2$.

### 2.7 Validation studies

We have performed calculations on a Catphan phantom and a head-and-neck (HN) cancer patient to demonstrate the feasibility of using our gDRR package for computing realistic CBCT projection images. Meanwhile, the computational efficiency is assessed by recording the computation time at each key step. For the hardware used in this section, the GPU is an NVIDIA GTX580 card equipped with 512 processors and 1.5 GB GDDR5 memory. A desktop computer with a 2.27 GHz Intel Xeon processor and 4GB memory is also used, on which EGSnrc code is executed for the purpose of validating our MC simulations.

Our computations are conducted under a geometry resembling that of the kV OBI system on a TrueBeam linear accelerator. The x-ray source-to-axis distance is 100 cm and the source-to-detector distance is 150 cm. The x-ray imager resolution is $512 \times 384$ with a pixel size of $0.776 \times 0.776$ mm². The detector is positioned in the opposite direction to the source, namely $\theta = 0°$ in Figure 1, and the point $P$ is at the center of the detector for a full-fan scan setup. For the purpose of demonstrating principles, the x-ray source energy spectrum, the detector response, and the source photon fluence map are chosen as those shown in Figure 2. For the phantom size, the Catphan phantom is used for calibration with a resolution of $256 \times 256$ voxels in a transverse slice with the in-plane voxel size chosen to be $1.0 \times 1.0$ mm². As for the HN patient, its in-plane resolution is $512 \times 512$ voxels with a voxel size chosen to be $0.976 \times 0.976$ mm². Both phantoms have the slice thickness of 2.5 mm with the number of slices large enough to cover the phantoms. $5 \times 10^8$ source photons are simulated for each projection image in MC simulations, unless stated otherwise.

## 3 Results

### 3.1 Primary signal

The first result we present is the primary signal of the projection image for a HN patient with the CBCT source on the left side. Figure 4(a) shows the ray-tracing result with the tri-linear interpolation algorithm. We have also presented the primary signals computed by the MC simulation with density interpolation in Figure 4(b). These two figures are visually close to each other, although a certain amount of noise presents in the MC results. Artifacts caused by the finite voxel size are still observable, especially in the zoom-in view. Figure 5 plots the profiles along the coordinate axes $u$ and $v$ shown in Figure 2(c) of the primary signal $P$ computed from the tri-linear ray-tracing algorithm and $\hat{P}$ from the MC simulations with density interpolation switched on. These two signals agree well. Taking the difference between $P$ and $\hat{P}$ yields the noise $\hat{N}_P$. Note that the amplitude of $\hat{N}_P$ is governed by the number of photons simulated in the MC simulation, but do not represent the real noise level in an experiment.

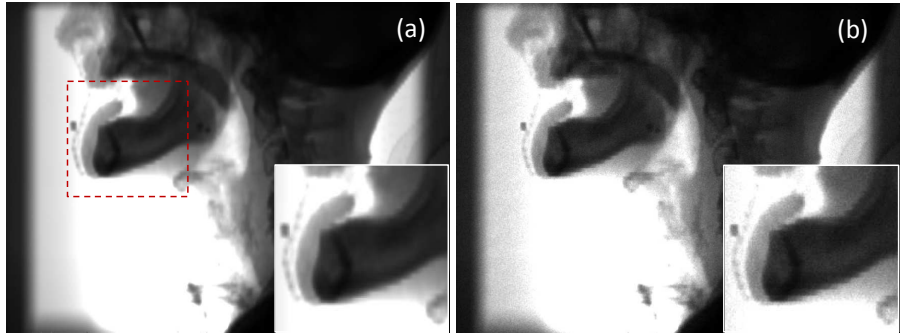

Figure 4. Primary signal simulated in gDRR of a HN patient by ray-tracing method using the tri-linear interpolation algorithm (a) and MC simulations with density interpolation (b). Inserts show a zoomed-in view of the area indicated by the square in (a).

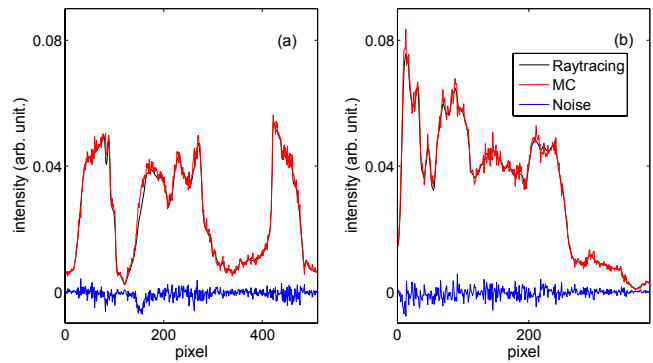

Figure 5. Intensity profiles along the $u$ and the $v$ axes of the primary signal $P$ computed from tri-linear ray-tracing and $\hat{P}$ from MC simulations with density interpolation switched on. The difference between them is the unscaled noise $\hat{N}_P$.

## 3.2  Scatter signal

For the scatter signal calculations, we first show in Figure 6(a) and (b) the signals calculated for a HN patient along the left-right projection direction with $5 \times 10^8$ and $5 \times 10^{10}$ source photons, respectively. There is a visible amount of noise in Figure 6(a). The noise level is diminishing when the source photon number becomes large in Figure 6(b), as expected. In Figure 6(c), the scatter image after removing the noise component from Figure 6(a) is presented. Visually, Figure 6(b) and (c) are very close to each other.

We have also quantitatively computed the relative difference between these two images, *e.g.* $\|\hat{S}_{5\times10^{10}} - S_{5\times10^8}\|_2 / \|\hat{S}_{5\times10^{10}}\|_2$, where $\hat{S}_{5\times10^{10}}$ is the MC simulated image in (b) and $S_{5\times10^8}$ is the denoised image in (c). $\|.\|_2$ stands for the l-2 norm. This relative difference is about 2%, indicating the closeness of our denoised image to the underlying ground truth, which is considered to be the MC result with a very large number of photons in (b). The small, but finite, relative difference can be ascribed into the residual noise component in Figure 6(b), though it is hardly visible in the image.

We further plot the scatter image profiles along the $u$ and the $v$ axes in Figure 6(d) and (e), as well as the difference between the MC result $\hat{S}$ and the denoised result $S$, namely $\widehat{N}_S = \hat{S} - S$. Finally, in Figure 6(f) we depict the profiles of the denoised image based on simulations with a wide range of number of photons. The image profiles from $5\times10^7$ to $5\times10^{10}$ photons almost coincide on a single curve, and only in the case with $5\times10^6$ photons do we observe a slight difference. This indicates that the denoising algorithm is robust, such that the scatter signal can be estimated with a much-reduced number of photons.

To test the accuracy of the simulated scatter components, we have validated our simulations against EGSnrc[9, 10], a commonly used MC package for photon transport. For this purpose, the scatter image under the identical configuration as in the above case is performed with EGSnrc. $10^9$ source photons are used in the simulation. Due to the different signal intensities from EGSnrc and from gDRR, we rescale the EGSnrc result, so that its mean value equals to that of the gDRR results. The resulting scatter signal and the corresponding denoised image are shown in Figure 7(a) and (b), respectively. Visually, these two images are indistinguishable from the corresponding results in gDRR, namely Figure 6(a) and (c). We have also plotted the denoised image profiles along the $u$ axis for the two simulation packages. A good agreement is observed in Figure 7(c). Finally, we calculated the relative difference of the scattering signals $\frac{\|S_{gDRR} - S_{EGSnrc}\|_2}{\|S_{EGSnrc}\|_2} = 3.8\%$. This value quantitatively indicates the accuracy of the scatter simulations in gDRR.

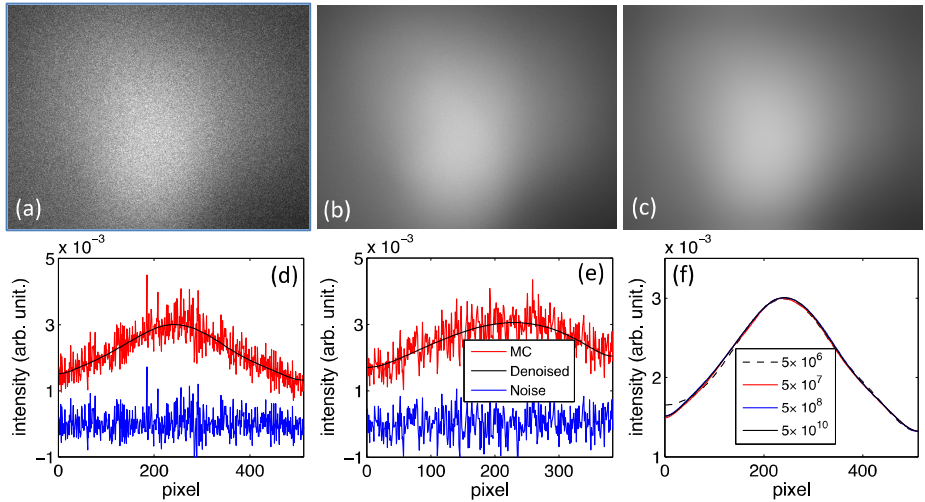

Figure 6. (a) and (b) are scatter signals of a HN patient simulated using gDRR with $5\times10^8$ and $5\times10^{10}$ source photons. (c) shows the denoised scatter image from simulations in (a). (d) and (e) show the signal profiles along the $u$ and the $v$ axes, respectively. (f) is the comparisons of the denoised scatter image profiles along the $u$ axis with various number of photons.

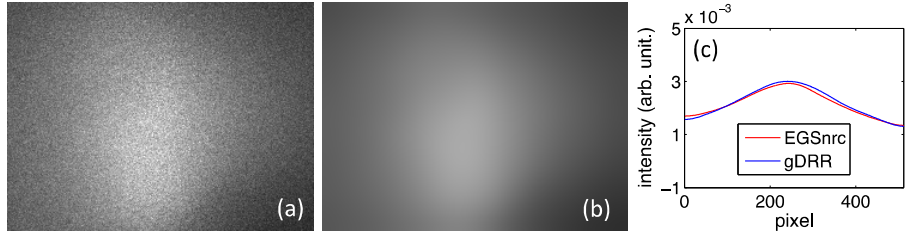

Figure 7. (a) Scatter signal of the HN case using EGSnrc. (b) Denoised result of (a). (c) A comparison of denoised scatter image profiles along the $u$ axis between the EGSnrc result and the gDRR result.

*3.3 Noise signal*

The noise component calibration as described in the section 2.6 is conducted. For a Catphan phantom at the homogeneous layer as indicated in Figure 8(a), we first calculated the quantity $\sigma_X(u, I)/\sigma_{\widehat{N}}(u)$ at a given mAs level $I$ and plotted the results as a function of $u$ in Figure 8(b). Apart from the noise, this quantity is almost a constant independent of the pixel location $u$, which indicates the validity of our noise model. Moreover, we average $\sigma_X(u, I)/\sigma_{\tilde{N}}(u)$ for a range of pixel locations $u$ for a given level of mAs and then plot the averaged $\langle \sigma_X(u, I)/\sigma_{\tilde{N}}(u) \rangle$ as a function of $1/\sqrt{I}$ in Figure 8(c). The resulting data are found to be along a straight line. A linear regression yields the slope $k = 0.3074$ and hence the level of $\zeta$ can be derived as $\zeta = n_{\text{sim}}/k^2 = 5.2913 \times 10^9$. This value enables us to convert the simulated noise signal into the realistic noise signal according to the given number of photons and the desired mAs level based on the Eq. (6).

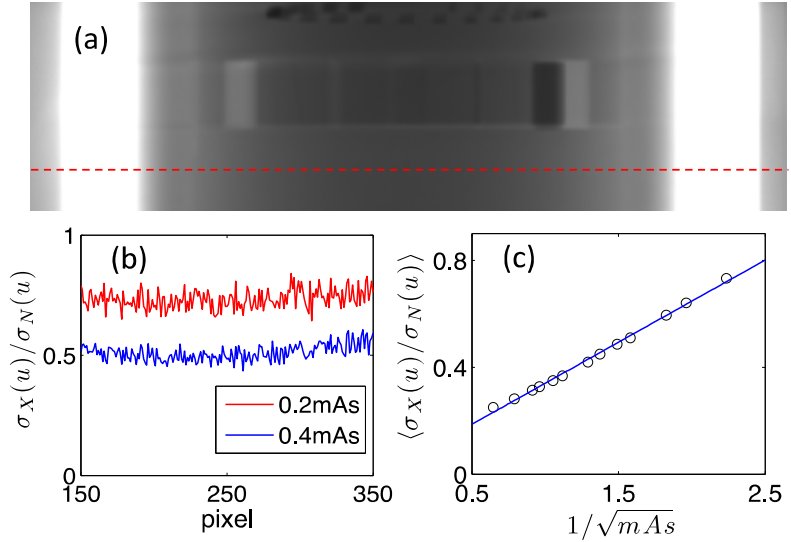

Figure 8. (a) One projection of the Catphan phantom. Red line indicates the location used for noise calibration. (b) The values of $\sigma_X(u, I)/\sigma_{\widehat{N}}(u)$ as a function of $u$ at two different mAs levels. (c) $\langle \sigma_X(u, I)/\sigma_{\widehat{N}}(u) \rangle$ as a function of $1/\sqrt{mAs}$ and the blue line is the linear fit.

## 3.4 Computation time

To assess the computational efficiency, we have recorded the computation time of each key step in both the Catphan phantom case and the HN patient case. The results are summarized in Table 1. All computation time are expressed as per projection. Note that a poly-energetic spectrum with about 100 energy channels is used in these cases. Hence the ray-tracing calculation time is longer than what has been previously reported in other similar research works[40, 41]. The ray-tracing time and the MC simulation time for the HN patient case are larger than those for the Catphan phantom case, due to the large number of voxels in the former. On the other hand, the computation time for the denoising and noise calibration part is independent of the phantoms, as both tasks operate in the projection image domain. Among all of the steps in gDRR, MC simulation is the most time-consuming. Comparing with the ray-tracing calculations and the MC simulations, the time spent on denoising and noise calibration tasks is negligible. Overall, these recorded time clearly indicate the achieved high efficiency in gDRR. For instance, the computation time for EGSnrc for the results shown in Figure 7 is ~16 CPU hours.

Table 1. Computation time (in second) of each key step in gDRR.

|            | Ray-tracing | MC simulation | Denoising & Noise Calibration |
|------------|-------------|---------------|-------------------------------|
| Catphan    | 1.2         | 28.1          | 0.041                         |
| HN patient | 2.3         | 95.3          | 0.042                         |

## 3.5 CBCT artifacts

It is well known that various artifacts can be observed in realistic CBCT images due to various physical processes involved in data acquisition. To demonstrate the feasibility of using gDRR to compute realistic x-ray projection images, we have simulated 360 projection images of the HN patient in an angular range of $2\pi$ at 0.6 mAs/projection and have reconstructed the CBCT image using an FDK algorithm. Various artifacts are observed in the CBCT images reconstructed as such. First, Figure 9(a) and (b) show one slice of the CBCT with a mono-energetic 60 keV source and a poly-energetic 100 kVp source, respectively. Only primary signals are used in these two reconstructions.

Comparing these two images, artifacts caused by beam-hardening effect are clearly observed in (b), as indicated by the arrow. Figure 9(c) is the same CBCT slice but reconstructed with all of the primary, the scatter, and the noise signals. A poly-energetic 100 kVp source is used in this case. Apart from the obvious level of noise due to the inclusion of noise signal in the projections, scatter-caused artifacts are also observed, which reduce the overall image contrast, and strengthen the artifacts evidenced by the arrow. Finally, in Figure 9(d) we show another slice of the CBCT reconstructed in the same simulation setup as in (c), but at a different display window level. An obvious ring shadow artifact presents due to the interplay between the scatter signal and the bow-tie

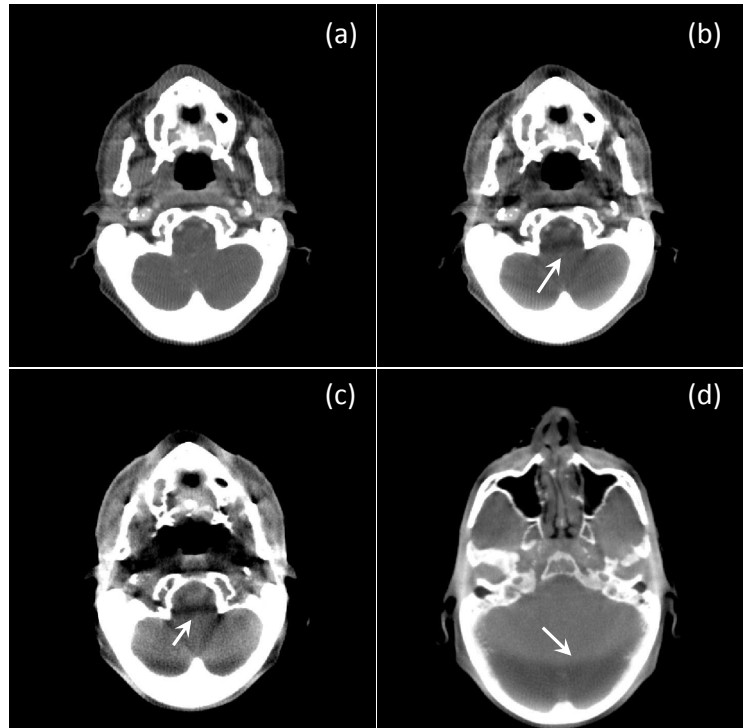

Figure 9. CBCT images reconstructed from the simulated projections. (a) From only primary signal with a mono-energetic source. (b) From primary signal with a poly-energetic source. (c) and (d) From all three components. The display window is [-100, 340] HU for (a), [20, 460] HU for (b) and (c), and [-400, 890] HU for (a). Arrows indicate various CBCT artifacts.

filter. This artifact can also be observed in Figure 9(c).

As a validation of the noise model, we have attempted to reconstruct the CBCT image of the Catphan phatom using the simulated projections at various mAs levels. A square region of interest (ROI) is selected in the center of the transverse CBCT slice inside the homogeneous phantom layer and noise amplitude in the reconstructed CBCT images $\sigma_{sim}$ are measured as the standard deviation inside the ROI. Meanwhile, such a phantom is scanned under the CBCT system with the same mAs levels and the CBCT images are reconstructed. Noise amplitude $\sigma_{exp}$ is also measured in the same ROI. The same FDK algorithm is used to reconstruct CBCT images in all cases for a fair comparison between the simulation and the experimental studies. Figure 10 plots the noise amplitudes $\sigma_{\text{sim}}$ and $\sigma_{\text{exp}}$ as functions of $1/\sqrt{mAs}$. The two function curves are in good agreement, indicating the capability of gDRR in terms of reproducing noise signals in the projection images and hence in the reconstructed CBCT images.

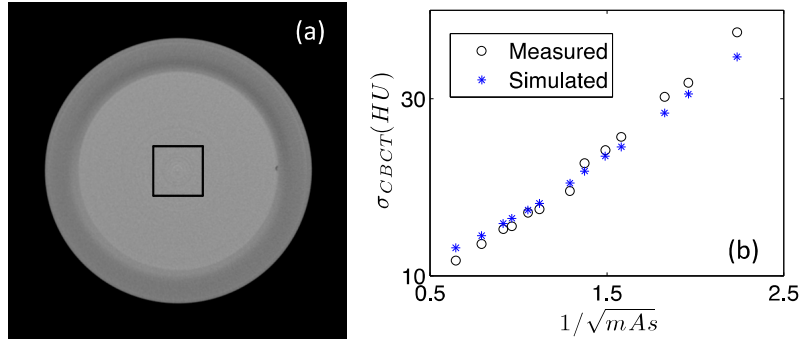

Figure 10. (a) The homogeneous layer of the Catphan phantom. The square indicates the ROIs selected for noise comparison. (b) Comparison of the noise amplitudes from the simulated results and from the experimental results.

## 4      Conclusion and Discussions

In this paper, we have presented our recent progress towards the development of a GPU-based package gDRR for the simulations of x-ray projections in CBCT with a number of realistic features included, *e.g.* a bowtie filter, a poly-energetic spectrum, and detector response. The input of gDRR includes a voxelized phantom data that defines material type and density at each voxel, x-ray projection geometry, as well as source and detector properties. gDRR then computes three components, namely primary, scatter, and noise at the detector. The primary signal is computed by a tri-linear ray-tracing algorithm. An MC simulation is then performed, yielding the primary component and the scatter component, both with noise. A denoising process specifically designed for Poisson noise removal is applied to generate the smooth scatter signal. The noise component is then obtained by taking the sum of the difference between the MC primary and the ray-tracing primary, and the difference between the MC simulated scatter and the denoised scatter. Finally, a calibration stage converts the calculated noise to a realistic noise by scaling its amplitude according to the desired mAs levels. The calculated projections are found to be realistic, such that various artifacts in real CBCT images can be reproduced by the simulated projections, including beam hardening, scattering, and noise levels. gDRR is developed on the GPU platform with a finely tuned structure and implementations to achieve a high computational efficiency.

Although gDRR is initially developed and calibrated for the OBI system on a TrueBeam machine, with simple modifications, it can, with simple modifications, also be applied to the simulations of x-ray projection images in other geometry, such as C-arm

CBCT. Also, each components of the package, namely the calculations of the primary and the scatter signals, can be singled out for different research purposes. The scatter signals simulations can also be configured to tally scatter photons of different types and orders. These features greatly enable the wide applicability of gDRR and facilitate CBCT-related research projects in a variety of contexts. The entire package will be in public domain for research use, and is currently available upon request.

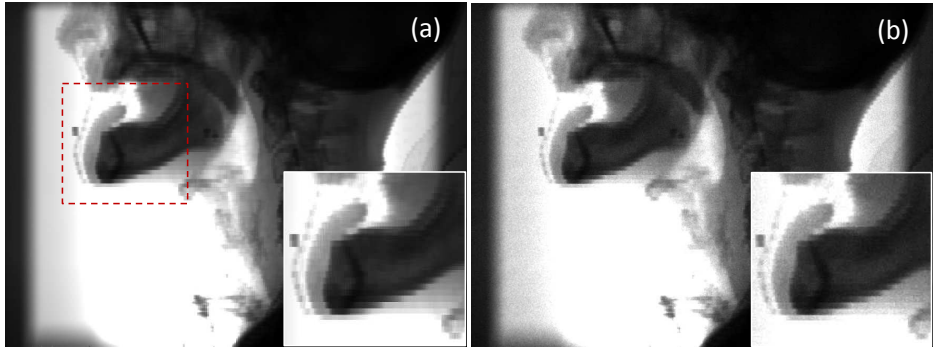

Figure 11. Primary signal simulated in gDRR of a HN patient by ray-tracing method using the Siddon's algorithm (a) and MC simulations without density interpolation (b). Inserts show a zoomed-in view of the area indicated by the square in (a).

Siddon's ray-tracing algorithm is also available in gDRR for primary signal calculation, although it is not the default algorithm for this purpose. A user can select this option, if the block artifacts are not a concern and a higher computational efficiency is more desired. Figure 11(a) shows the primary projection signal for the same patient case as in Figure 4. When comparing Figure 11(a) and Figure 4(a), especially the inserts, it is found that the projection computed by the Siddon's algorithm has more apparent artifacts caused by the finite voxel size. On the other hand, because of the absence of tri-linear interpolation and hence the reduced memory access, the Siddon's algorithm attains a higher computational efficiency, such that the computation time for the result in Figure 11(a) is 1.9 sec, a 17% percent improvement compared to the time of 2.3 sec reported in the Table 1. Similarly, when the density interpolation is switched off in the MC simulations, the finite voxel size causes more obvious artifacts in the primary projection image, as illustrated in Figure 11(b). The computation time is, however, shortened from 95.3 sec to 80.7 sec. The impacts of density interpolation on the scatter signal calculation are not observed.

**Acknowledgements**

This work is supported in part by NIH (1R01CA154747-01), the University of California Lab Fees Research Program, the Master Research Agreement from Varian Medical Systems, Inc., and the Early Career Award from the Thrasher Research Fund.